\documentclass[journal=jacsat,manuscript=reprint]{achemso}

\usepackage{chemformula} 
\usepackage[T1]{fontenc} 
\setkeys{acs}{maxauthors = 0}
\author{Marianna Gerina}
\affiliation[CUNI]
{Department of Inorganic Chemistry, Charles University, Hlavova 2030/8, 128 43 Prague 2, Czech Republic}
\author{Marco Sanna Angotzi}
\affiliation[UNICA]
{Department of Chemical and Geological Sciences, University of Cagliari, S.S. 554 bivio per Sestu, 09042 8 Monserrato (CA), Italy}
\author{Valentina Mameli}
\affiliation[UNICA]
{Department of Chemical and Geological Sciences, University of Cagliari, S.S. 554 bivio per Sestu, 09042 8 Monserrato (CA), Italy}
\author{Michal Mazur}
\affiliation[CUNI2]
{Department of Physical and Macromolecular Chemistry, Charles University, Hlavova 2030/8, 128 43 Prague 2, Czech Republic}
\author{Nicoletta Rusta}
\affiliation[UNICA]
{Department of Chemical and Geological Sciences, University of Cagliari, S.S. 554 bivio per Sestu, 09042 8 Monserrato (CA), Italy}
\author{Elena Balica}
\affiliation[Firenze]
{Dipartimento di Chimica and INSTM, Universita`di Firenze, Via della Lastruccia 3, I- 50019 Sesto Fiorentino, Italy}
\author{Pavol Hrubovcak}
\affiliation[UJP]{Institute of Physics, Faculty of Science, P.J. Šafárik University, Park Angelinum 9, 04001 Košice, Slovakia}
\author{Carla Cannas}
\affiliation[UNICA]
{Department of Chemical and Geological Sciences, University of Cagliari, S.S. 554 bivio per Sestu, 09042 8 Monserrato (CA), Italy}
\author{Dirk Honecker}
\affiliation[ISIS]{ISIS Neutron and Muon Facility, Science and Technology Facilities Council,Rutherford Appleton Laboratory, OX11 0QX Didcot, United Kingdom}
\author{Dominika Z\'akutn\'a}
\affiliation[CUNI]
{Department of Inorganic Chemistry, Charles University, Hlavova 2030/8, 128 43 Prague 2, Czech Republic}
\email{zakutnad@natur.cuni.cz}

\title[An \textsf{achemso} demo]
  {Exploring Anisotropy Contributions in Mn$_\mathrm{x}$Co$_\mathrm{1-x}$Fe$_2$O$_4$ Ferrite Nanoparticles for Biomedical Applications}

 
\begin{document}


\begin{abstract}
  Designing well-defined magnetic nanomaterials is crucial for various applications and it demands a comprehensive understanding of their magnetic properties at the microscopic level. In this study, we investigate the contributions to the total anisotropy of Mn-Co mixed spinel nanoparticles. By employing neutron measurements sensitive to the spatially resolved surface anisotropy with sub-\AA\space resolution, we reveal an additional contribution to the anisotropy constant arising from shape anisotropy and interparticle interactions. Our findings shed light on the intricate interplay between chemical composition, microstructure, morphology, and surface effects, providing valuable insights for the design of advanced magnetic nanomaterials for AC biomedical applications, such as cancer treatment by magnetic fluid hyperthermia.
\end{abstract}

\section{Introduction}
\begin{figure}[htbp!] 
	\begin{center}
		\advance\leftskip-0.4cm
		\advance\rightskip0cm
		\includegraphics[width=1\columnwidth]{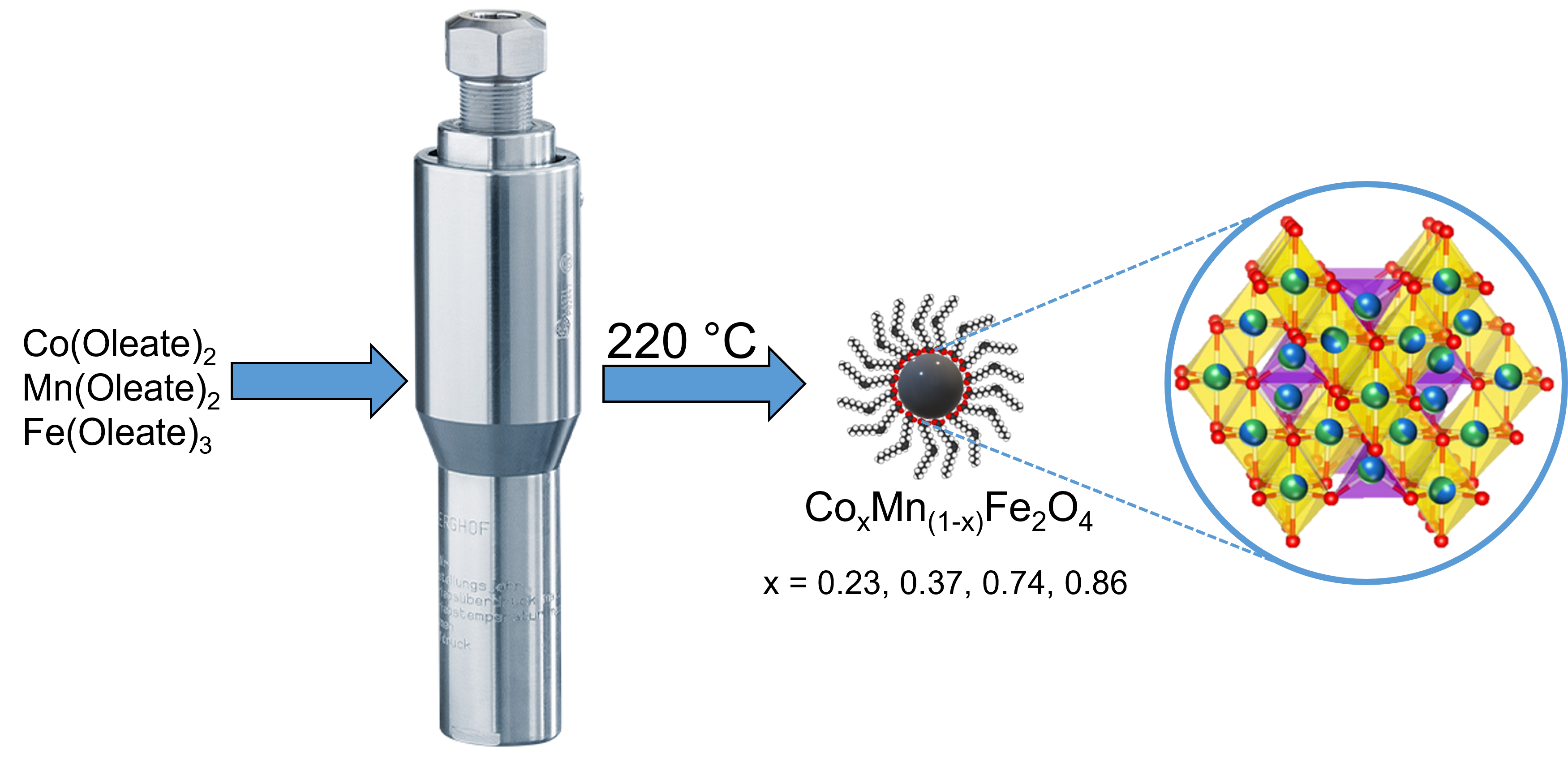}
		\caption{Scheme of the solvothermal synthesis of Co$_\mathrm{x}$Mn$_\mathrm{1-x}$Fe$_2$O$_4$.}
		\label{fig:Synthesis}
	\end{center}
\end{figure}

The magnetic properties of spinel ferrite nanoparticles (NPs) result from a multifaceted interaction of factors such as chemical composition, microstructure, morphology, and surface phenomena \cite{MUSCAS2018127}. Among the spinel ferrites, cobalt ferrite stands out by exhibiting a magnetically hard behavior, which is attributed to the high single-ion anisotropy of cobalt ions. Since the 2000s, strategies for fine-tuning its magnetic behavior and reducing its toxicity through cation substitution (Mn, Zn, etc.) have been suggested \cite{Hochepied_2000}.  However, establishing a correlation between the magnetic properties of NPs and their performance can pose a challenge due to the simultaneous occurrence of diverse structural and magnetic phenomena.
A suite of techniques is typically employed to elucidate the interconnections between the structural and magnetic properties of NPs. These include transmission electron microscopy (TEM), dynamic light scattering (DLS), small-angle X-ray scattering (SAXS), DC magnetization, and AC susceptibility (ACS) measurements \cite{Bender_2017}. The theoretical framework outlines the relationships between particle structure and magnetic properties in the ideal scenario of homogeneously magnetized, single-domain, and non-interacting particles. However, recent findings underscore that NP systems are more intricate than previously assumed, even for highly crystalline, monodisperse, and non-interacting NP ensembles. Indeed, several factors can affect the material's magnetic properties, such as spin disorder and dipolar interactions.
Spin disorder, \textit{i.e.}, random orientation of magnetic spins,  is an inherent characteristic of magnetic NPs, notably when their diameter is reduced to a few nanometers \cite{Mørup2013}. The phenomena of spin canting (defined as a partial alignment of the spins at the surface under high magnetic fields, \textit{i.e.}, noncollinear magnetic spin structures) and spin frustration (presence of surface atoms with a reduced number of magnetic neighbor ions around) are frequently analyzed in the context of surface effects, attributed to the unsaturated covalent bonds of surface atoms. This perspective presumes that the entire NP is crystallographically perfect, with only the surface atoms \cite{Mørup2013}exhibiting distinct magnetic properties due to symmetry disruption. However, in actual samples, the atomic structural arrangement near the surface is considerably more intricate, with a multitude of lattice-level perturbations such as point and line defects, local strains, variations in the degree of inversion in the spinel structure, and cation depletion, to cite a few. These factors may result in disordered magnetic phases or a disordered shell of a few nanometers thicknesses \cite{Mørup2013,Gerina2023}, the extent of which is dependent on the chemical composition and synthesis method \cite{Aurelio2020}.
For instance, in the case of highly crystalline cobalt ferrite NPs devoid of internal defects, the spin disorder permeates the entire particle, resulting in the formation of a uniformly disordered phase \cite{Miksatko_2019}. These observations imply that the spin disorder at the surface triggers a comprehensive spin reorganization within the particle. The presence of cobalt tends to favor the extended disordered configuration, particularly for larger NPs. At the same time, magnetically soft ferrites are prone to form more diverse spin configurations, even though the magnetic core-shell structure is one of the prevalent ones \cite{Pacakova_2017}. A type of magnetic core-shell structure, primarily originating from the structural disorder, has been reported for cobalt ferrite NPs \cite{Aurelio2020}. Moreover, additional spin canting arises from intra- or inter-particle dipolar interactions \cite{Oberdick2018}, known, for instance, to alter the heating efficiency of NPs. These dipolar interactions, leading to particle agglomeration, introduce an element of uncertainty in the determination of the effective anisotropy constant ($K_\mathrm{eff}$) due to the lack of experimental control over the colloidal state of individual magnetic NPs during the measurement process \cite{Bender_2017}. This factor could account for discrepancies between the ferrofluid results and those derived from NPs supported in solid matrices or powdered preparations \cite{García-Acevedo}. As observed, disorder effects, being inherent to the materials themselves, play a crucial role in determining their magnetization properties, including coercivity, superparamagnetism, exchange interactions, and spontaneous magnetization \cite{Salazar_Alvarez_2008}. These properties are of paramount importance for the wide range of technological applications of magnetic NPs, such as in batteries\cite{Li2015}, biomedicine\cite{Mohsin2022}, or catalysis\cite{Zhang2019}. Surface spin canting or disorder in magnetic NPs can only be indirectly accessed using magnetization measurements, ferromagnetic resonance, Mössbauer spectroscopy \cite{PIANCIOLA201544}, X-ray magnetic circular dichroism \cite{Bonanni2018}, and electron energy loss spectroscopy \cite{Negi2017}. There are probes that can provide essential information on the local distribution of the magnetic moment, such as electron holography\cite{Pichon2011}, and electron magnetic circular dichroism\cite{Bueno2021}. Magnetic small-angle neutron scattering (SANS) is a versatile technique to directly obtain spatially sensitive information about the spin structure in NPs on the relevant nanometer length scale from the whole sample volume. SANS was utilized as a methodological approach to examine the interaction within ensembles of NPs \cite{Bender2018}, to probe the response of magnetic colloids \cite{Bender2015} and ferrofluids \cite{Gazeau_2002} when subjected to external fields. The application of half-polarized SANS (SANSPOL) facilitates the resolution of the quantitative distribution of magnetization within NPs. This confirmed the existence of spin disorder at the particle's surface while concurrently uncovering a substantial degree of spin disorder pervading the entire NP \cite{ZakutnaPhysRevX2020}. Recently, using the oleate-based solvothermal method (see Fig.\,\ref{fig:Synthesis}), we synthesized highly crystalline spheroidal Mn$_\mathrm{x}$Co$_\mathrm{1-x}$Fe$_\mathrm{2}$O$_\mathrm{4}$ NPs, having Co/Mn ratio 6.1, 2.8, 0.6, and 0.3 for Co86Mn14, Co74Mn26, Co37Mn63, and Co23Mn77, respectively, and narrow size distribution (dispersity below $20\%$) and studied their macroscopic magnetic properties as a function of the chemical composition\cite{SannaAngotzi2021}. 

\begin{figure}[htbp!] 
	\begin{center}
		\advance\leftskip-0.4cm
		\advance\rightskip0cm
		\includegraphics[width=1\columnwidth]{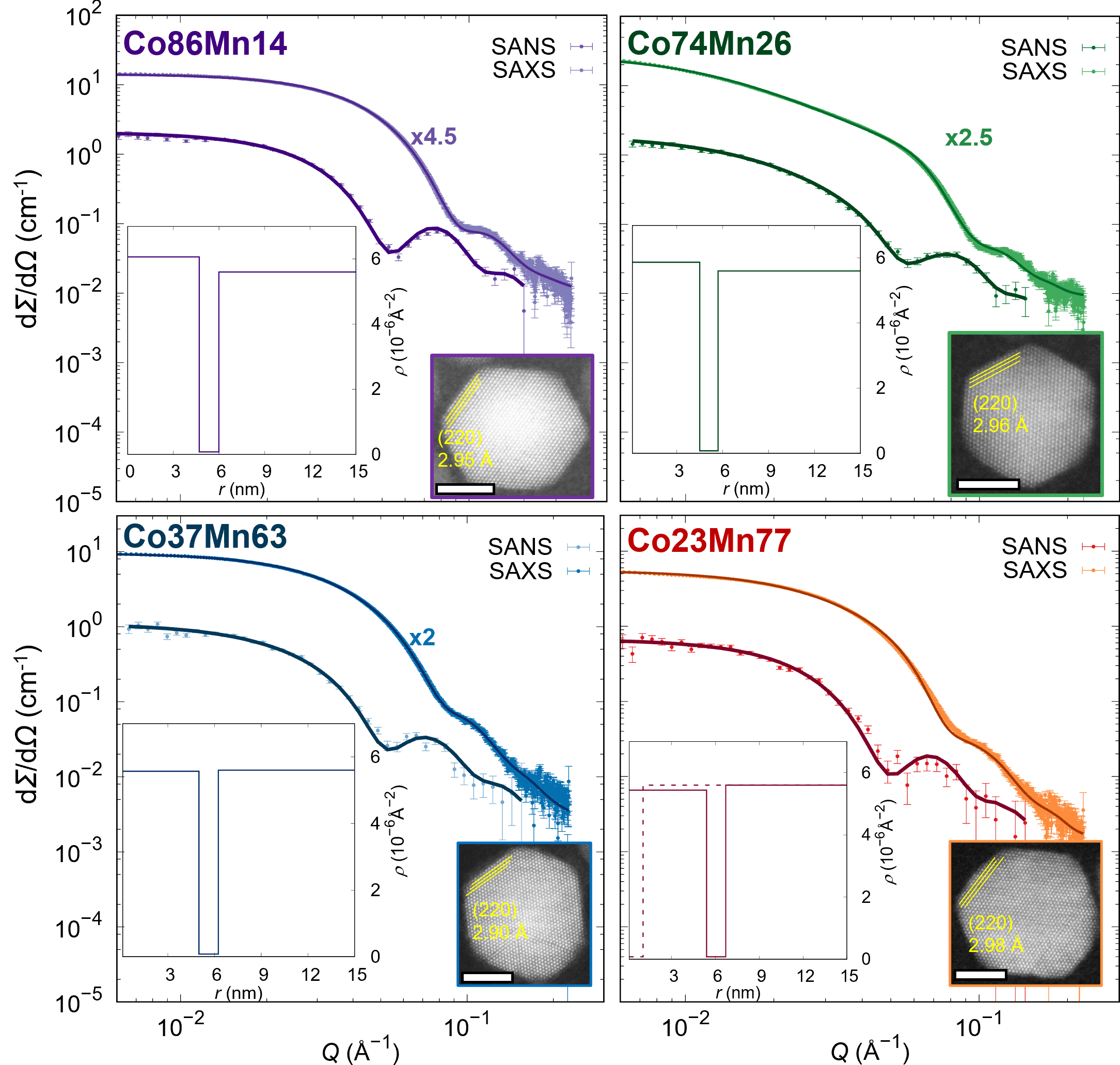}
		\caption{SAXS and pure nuclear SANS cross-section (points) with form factor refinements (full lines). The SAXS data were scaled by 4.5, 2.5, and 2 times, as indicated in the graphs above the SAXS data. Insets: left bottom corner - obtained radial distribution of $\rho_\mathrm{mag}$ SANS refinements. Continuous lines refer to the particle, and dashed lines correspond to extra free oleic acid micelles. Right bottom corner - HRSTEM micrograph. (Scale bars: 5\,nm)}
		\label{fig:SAXS_SANS_SLD}
	\end{center}
\end{figure}

We observed a decrease in the anisotropy (see Table S1) with increasing Mn content and its resulting influence on \textit{T$_\mathrm{max}$}, \textit{T$_\mathrm{diff}$}, \textit{T$_\mathrm{b}$}, and \textit{H$_\mathrm{c}$} (maximum and furcation point of the ZFC curve, blocking temperature, and coercivity, respectively)\cite{SannaAngotzi2021}. All the samples showed a deviation from the ideal case of non-interacting NPs observed in the IRM-DCD protocols as negative $\mathrm{\Delta}M$ peaks in the Co-richer samples (dominant dipolar interactions) and positive $\mathrm{\Delta}M$ for the Mn-richest sample (probably due to unmasked cubic anisotropy as a result of weaker dipolar interactions). Nevertheless, discerning the cause of the non-ideal behavior in our samples was not possible with the macroscopic techniques used in the previous work, thus, in this work, we further investigated them through SANSPOL. The measurements were conducted on the colloidal dispersions to limit the role of interparticle interactions. Therefore, the surface anisotropy constant and the magnetocrystalline anisotropy were obtained and correlated with the Mn content.
The combined refinements of the SAXS and nuclear SANS scattering cross-sections (Fig.\,\ref{fig:SAXS_SANS_SLD}) revealed the spheroidal morphology of Mn-mixed cobalt ferrite NPs. For both X-rays and neutrons, the scattering length densities of the inorganic cores were calculated from structural information obtained from Rietveld refinement and elemental content from the ICP experiments, previously presented in the paper of Sanna Angotzi \textit{et al.}\cite{SannaAngotzi2021}. The scattering length densities of the cores, oleic acid (OA), and solvent (toluene and d$_8$-toluene for X-rays and neutrons, respectively) were kept fixed during the refinement of the nuclear structure, and only the size of the inorganic particle, polydispersity and length of OA were refined. The SAXS data were fitted using the spherical form factor and at the first glance, except for the Co74Mn26 sample, they show the presence of the intensity plateau known as the Guinier plateau\cite{Zakutna2024}, representing a non-interacting NP system. In the Co74Mn26 sample, the non-negligible interparticle interactions are visible as an increase in the scattering intensity in the \textit{Q}-range of 0.02 - 0.05\,\AA$^{-1}$. Nevertheless, after detailed analysis only the sample Co86Mn14 did not show interparticle interactions. On the remaining samples, the structure factor "sticky-hard sphere" was included, suggesting the presence of non-negligible, although weak, particles interactions. Thus, the samples were centrifuged at 6500 rpm to separate the NPs agglomerates before the SANS analysis. The SANS data were best described with the core-shell form factor, where the core corresponds to the inorganic NP size and the shell to the oleic acid (OA) molecule at the NP surface. In the case of the Co23Mn77 sample, the nuclear SANS scattering cross-section revealed a large background contribution arising from the presence of free OA micelles in the dispersion and, thus, their form factor was included in the refinement (their nuclear scattering length density is depicted as a dashed line). In all samples, the obtained thickness of OA surfactant was in the range of 1.2 - 1.3\,nm, which agrees with previous studies\cite{ZakutnaPhysRevX2020}. All samples have a size distribution below $20\%$ (Table \ref{table:Nuclear_dispersions}), representing minimal size effects on the NPs' magnetic response. Resolved nuclear morphology of Mn-mixed cobalt ferrite nanospheres was then fixed for the description of the SANS cross-section with various incident beam polarizations ($I_\mathrm{Q}^{-}$,$I_\mathrm{Q}^{+}$).

\begin{table*}[t]
	\centering
	\begin{tabular}{ccccccc}
		\hline
		\hline
		$\begin{array}{l} \textbf{Parameter} \\\hline \textbf{Sample} \end{array}$ & \textbf{Composition} & $d_\mathrm{nuc}$ & $d_\mathrm{OA}$ & $\rho_\mathrm{n}$ & $\sigma$\\
		& & (nm) & (nm) & (10$^{-6}$\,\AA$^{-2}$) & (\%)\\
		\hline
		\textbf{Co86Mn14} & Co$_\mathrm{0.77}$Mn$_\mathrm{0.13}$Fe$_\mathrm{2.07}$O$_\mathrm{4}$ & 9.4(1) & 1.31(1) & 6.062 & 14.9(1)\\
		\textbf{Co74Mn26} & Co$_\mathrm{0.66}$Mn$_\mathrm{0.23}$Fe$_\mathrm{2.07}$O$_\mathrm{4}$ & 9.0(1)  & 1.20(1) & 5.873 & 18.9(1)\\
		\textbf{Co37Mn63} & Co$_\mathrm{0.31}$Mn$_\mathrm{0.52}$Fe$_\mathrm{2.11}$O$_\mathrm{4}$ & 10.0(1) & 1.25(1) & 5.572 & 17.8(1)\\
		\textbf{Co23Mn77} & Co$_\mathrm{0.19}$Mn$_\mathrm{0.65}$Fe$_\mathrm{2.11}$O$_\mathrm{4}$ & 10.8(1) & 1.31(1) & 5.439 & 16.7(1)\\
		\hline
		\hline
	\end{tabular}
	\caption{Summarized results from SAXS and nuclear SANS refinements, where $d_\mathrm{nuc}$, $d_\mathrm{OA}$, $\rho_\mathrm{n}$, and $\sigma$ correspond to the nuclear particle diameter, the thickness of the capping agent, nuclear scattering length density of the core,  and lognormal size distribution, respectively.}
	\label{table:Nuclear_dispersions} 
\end{table*}

\begin{figure}[htbp!]
	\begin{center}
		\advance\leftskip-0.4cm
		\advance\rightskip0cm
		\includegraphics[width=1\columnwidth]{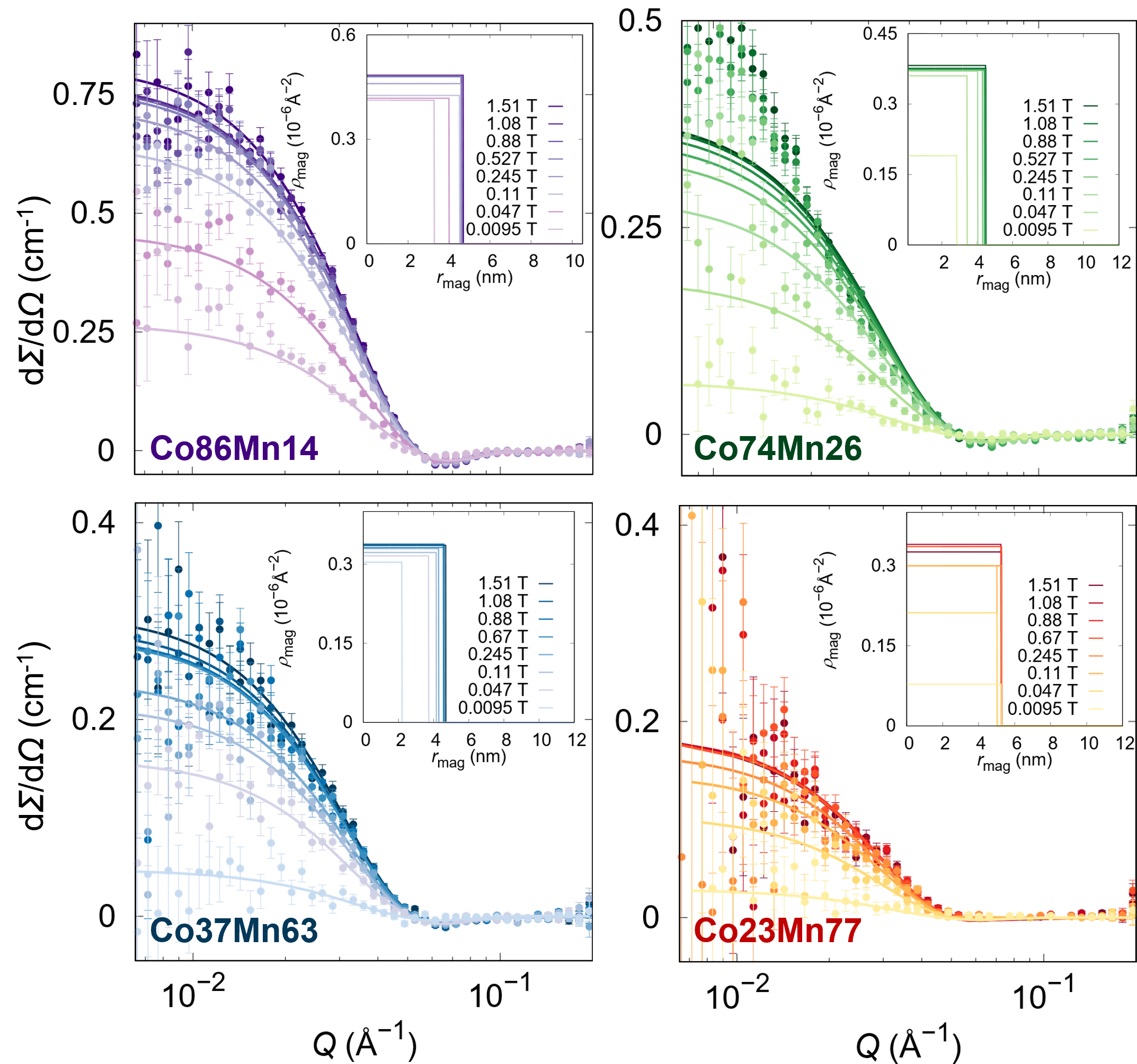}
		\caption{SANSPOL magnetic interference term (points) with core-shell-disorder layer form factor refinements (full lines) at different applied magnetic fields. Insets: Obtained radial distribution of $\rho_\mathrm{mag}$ at different applied magnetic fields.}
		\label{fig:Crossterm}
	\end{center}
\end{figure}

In our previous work\cite{SannaAngotzi2021}, the effective anisotropy constant obtained from magnetometry was found to decrease with the amount of Mn (see Table S1). To investigate more deeply the cause of the change in the effective anisotropy constant and the spin morphology of the NPs, we performed SANSPOL experiments. All samples show splitting of $I_\mathrm{Q}^{-}$ and $I_\mathrm{Q}^{+}$ scattering cross-sections arising from the core magnetic scattering (see Fig.\,S1 and Table\,S2). The data were described by the core-shell-surfactant model, where the whole physical inorganic NP size is fixed to the nuclear size obtained from SAXS and nuclear SANS, the shell represents the inorganic part with no magnetic scattering contribution that arises from spin disorder or canting, \textit{i.e.}, a disorder layer ($r_\mathrm{nuc} = r_\mathrm{mag} + d_\mathrm{dis}$), and the surfactant is the OA on the surface. To ensure no contribution from misaligned spins, we also performed the refinement of the nuclear-magnetic interference term (Fig.\,\ref{fig:Crossterm} and obtained results are summarized in Table S3), by subtracting $I_\mathrm{Q}^{-}$ from $I_\mathrm{Q}^{+}$, leading to similar observations. In fact, the samples Co89Mn14, Co74Mn26, and Co23Mn77 show the same behavior. Conversely, the sample Co37Mn63 shows the presence of a disorder layer and a change in $\rho_\mathrm{mag}$ similar to the Co-richest sample. The only refined parameters were the magnetic scattering length density of the core and the shell thickness without magnetic contribution. The radial distribution of magnetic scattering length density reported in the insets of Fig.\,\ref{fig:Crossterm} shows that the Co23Mn77 sample has a negligible change in the non-magnetized shell thickness, while the magnetized volume of the sample Co86Mn14, Co74Mn26, and Co37Mn63, richer in cobalt, is slightly increasing with an applied magnetic field. This could suggest the correlation between the presence of surface spin disorder and the amount of Co, which might also be related to a different bond strength between oleate molecules and other cations (Mn$^{2+}$, Co$^{2+}$, Fe$^{3+}$). 

The longitudinal magnetization, $M_\mathrm{z}(H)$, of all samples was accessed by refining the field dependence of magnetic SANS cross-sections and nuclear-magnetic interference term. The decrease in the longitudinal magnetization with Mn content in the sample was observed, which agrees with our already published observations from macroscopic magnetization measurements \cite{SannaAngotzi2021}. Moreover, the magnetic anisotropy was found to change throughout the series due to different cobalt content and the interplay between anisotropy and interparticle interactions. In fact, as a result of IRM-DCD protocols, except for the sample Co23Mn77, the $\mathrm{\Delta}M$ plots showed negative values of $\mathrm{\Delta}M$ and an upward concave curve in the Henkels plot as an indication of demagnetizing dipolar interaction or inhomogeneous spin structure \cite{DeToro2017}. The probability of having crystalline coherence between particles that would give rise to exchange interactions between particles was found to be very low in a previous study\cite{Frandsen2011}, favoring inhomogeneous spin structure. On the contrary, the sample Co23Mn77 showed positive values of $\mathrm{\Delta}M$ and a downward concave curve, which in the literature are commonly explained as due to magnetizing interactions, or cubic anisotropy \cite{Laureti_2010}.
From the macroscopic magnetic properties analysis, it was impossible to determine the cause of the observed behavior of the samples. Here, using SANSPOL, we can follow the change of magnetized volume for each sample, giving access to the possible surface spin disorder contribution. The results from the SANSPOL nuclear interference term show that the samples Co86Mn14, Co74Mn26, and Co37Mn63 exhibit a magnetically disordered layer on the surface. Combining these observations with the previous results obtained from $\mathrm{\Delta}M$ plots (see our previous results \cite{SannaAngotzi2021}), we could conclude that the negative values of $\mathrm{\Delta}M$ are due to the inhomogeneity of the spin structure at the surface.

From the $M_\mathrm{z}(H)$ and magnetized volume, the disorder energy can be calculated, according to Eq.\,1 in SI \cite{ZakutnaPhysRevX2020,Gerina2023}. From the derivative of the disorder energy with respect to the magnetized volume (Eq. 2 in SI), we can obtain the $K_\mathrm{eff-surf}$, the surface contribution to the total effective anisotropy. Table \ref{table:surfaceanisotropyconstant} reports that $K_\mathrm{eff-surf}$ decreases with increasing Mn content in the ferrite structure. Moreover, we calculated the surface anisotropy constant according to Z\'akutn\'a \textit{et al}. \cite{ZakutnaPhysRevX2020} (Eq. 3 in SI) and found that it is more significant for the Co-richest sample but stays constant for the other Co/Mn ratios.  

\begin{table*}[t]
	\centering
	\begin{tabular}{ccccc}
		\hline
		\hline
		\textbf{Parameter} & \textbf{Co86Mn14} & \textbf{Co74Mn26} & \textbf{Co37Mn63} & \textbf{Co23Mn77}  \\
		\hline
		$K$ (10$^{5}$ Jm$^{-3}$)\cite{SannaAngotzi2021} & 10.9(4) & 9.99(4) & 6.37(3) & 3.74(2) \\
		$K_\mathrm{eff-surf}$ (10$^{5}$ Jm$^{-3}$) & 5.4(3) & 4.0(5) & 3.5(5) & 3.5(9) \\
		$K_\mathrm{s}$ (mJm$^{-2}$) & 0.82(4) & 0.60(7) & 0.55(8) & 0.6(2) \\
		$K_\mathrm{dd-max}$ (10$^{5}$ Jm$^{-3}$) & 0.054 & 0.041 & 0.026 & 0.024 \\
		$K_\mathrm{shape}$ (10$^{5}$ Jm$^{-3}$) & 0.62(6) & 0.14(3) & 0.23(3) & 0.21(2) \\
		$K_\mathrm{A}$ (10$^{5}$ Jm$^{-3}$) & 4.8(5) & 5.8(5) & 2.7(5) & -0.01(1) \\
		\hline
		\hline
	\end{tabular}
	\caption{Obtained values of effective anisotropy $K$\cite{SannaAngotzi2021}, effective surface anisotropy $K_\mathrm{eff-surf}$, maximum surface anisotropy, $K_\mathrm{s}$, maximum dipole-dipole interaction contribution, $K_\mathrm{dd-max}$, shape anisotropy contribution, $K_\mathrm{shape}$, and remaining magnetocrystalline anisotropy, $K_\mathrm{A}$.}
	\label{table:surfaceanisotropyconstant} 
\end{table*}

\begin{figure}[H]
	\begin{center}
		\advance\leftskip-0.4cm
		\advance\rightskip0cm
		\includegraphics[width=1\columnwidth]{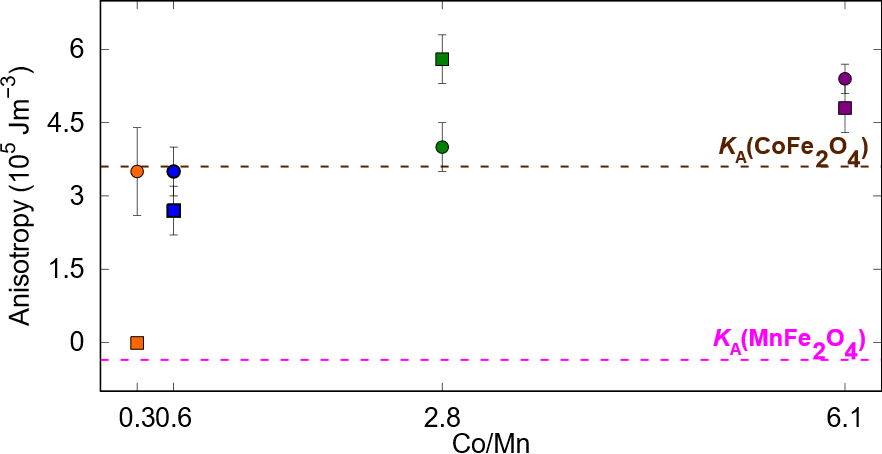}
		\caption{Trend of $K_\mathrm{eff-surf}$ (circles) and $K_\mathrm{A}$ (squares) as a function of the Co/Mn ratio. Pink and brown dashed lines indicate the bulk $K_\mathrm{A}$ for MnFe$_\mathrm{2}$O$_\mathrm{4}$ and CoFe$_\mathrm{2}$O$_\mathrm{4}$, respectively.}
		\label{fig:Anisotropycontributions}
	\end{center}
\end{figure}

Our previous study showed that surface anisotropy does not depend on the size of NPs\cite{Gerina2023}; moreover, here, we have clear confirmation that the chemical composition can also alter the surface anisotropy contribution, as already demonstrated in previous studies \cite{Aslibeiki2019}. 
Knowing the effective anisotropy constant from averaged magnetization measurements\cite{SannaAngotzi2021} and the averaged surface anisotropy contribution from SANSPOL allows us to extract the magnetocrystalline anisotropy constant and estimate the effect of Co/Mn content in the spinel structure. As shown in Fig.\,\ref{fig:Anisotropycontributions}, $K_\mathrm{eff-surf}$
increases with the Co/Mn ratio. Moreover, it is worth noting that the $K$ is obtained from powder measurements where non-negligible interparticle interactions might be present
in contrast to SANSPOL. In an earlier investigation of the magnetic interactions in Co-doped MnFe$_\mathrm{2}$O$_\mathrm{4}$, the dipolar interaction was observed in all the samples under study \cite{Aslibeiki2019}. Hence, we calculated the maximum dipole-dipole interaction contributions and compared them with the different anisotropy contributions. The contribution of interparticle interactions to the $K$ was estimated using the dipole-dipole interactions energy ($E_\mathrm{dd} = \mu_\mathrm{0}\mu^2/(4\pi r_\mathrm{ptp}^3$)). Considering the particle-to-particle distance, $r_\mathrm{ptp}$ in the closest proximity when the NPs are separated by an OA ligand leads to the values of interparticle anisotropy, $K_\mathrm{dd}$ of 10$^3$ order of magnitude (see Table \ref{table:surfaceanisotropyconstant}), which is negligible compared to surface anisotropy. An additional contribution to the total effective anisotropy arises from the shape of nanoparticles. Indeed, looking at HRSTEM micrographs (inset of Fig.\,\ref{fig:SAXS_SANS_SLD}, for more, see SI), the morphology of NPs is not perfectly spherical, but the majority of NPs are faceted. Taking into account the cuboidal shape of NPs (having (220) facets at the surface), the shape anisotropy constant can be expressed as $K_\mathrm{shape} = \frac{\mu_\mathrm{0}M^2_\mathrm{S}}{2}(N_\mathrm{zz} - N_\mathrm{xx})$, where $N_\mathrm{zz}, N_\mathrm{xx}$ are demagnetization tensor eigenvalues (for more information see SI). 

As we can see from HRSTEM micrographs, none of the NPs have a perfect spherical morphology but rather a cuboidal shape with a mean aspect ratio in the range of 1.06 - 1.13 (see SI). The obtained values of shape anisotropy (Table\,\ref{table:surfaceanisotropyconstant}) were subtracted from the total effective anisotropy, leading to the derived values of magnetocrystalline anisotropies ($K_A$).
As shown in Fig.\,\ref{fig:Anisotropycontributions}, the $K_A$ increases with the Co/Mn ratio. The reported room temperature bulk values of $K_A$ from pure \ch{CoFe2O4} and \ch{MnFe2O4} are 3.6$\cdot$10$^{5}$\,Jm$^{-3}$ and -0.36$\cdot$10$^{5}$\,Jm$^{-3}$, respectively\cite{Shenker,HOEKSTRA}. Looking at the obtained dependence of $K_A$ with the Co/Mn ratio, we can see that two samples (Co23Mn77 and Co37Mn26) are in the expected range of magnetocrystalline anisotropy. This result aligns with previously published results, where the anisotropy constant is observed to decrease with the amount of manganese due to the replacement of Co by Mn in the octahedral sites \cite{SannaAngotzi2021}. Co74Mn26 and Co86Mn14 samples have a high spinel inversion degree of 0.76 compared to the Co37Mn63 and Co23Mn77 samples where the inversion degree is significantly smaller with a value of 0.57, which also affects the magnetocrystalline anisotropy. However, in the case of the Co74Mn26 and Co86Mn14 samples, we have additionally noticed a large number of defects and grain boundaries in the crystal structure (see SI) that might further affect the magnetocrystalline anisotropy, which can lead to the enhancement of anisotropy associated with translational bond breaking, spin-orbit interaction, and enhancement of super-exchange interaction. 

In conclusion, a series of mixed Co/Mn ferrites NPs having Co/Mn molar ratios 6.1, 2.8, 0.6, and 0.3 and similar size and polydispersity were analyzed using SANSPOL to shed light on the behaviors observed from IRM-DCD protocols in our previous study. The samples with Co/Mn 0.6, 2.8 and 6.1 exhibited similar responses showing negative $\mathrm{\Delta}M$ values, while the Mn-richest sample (Co/Mn of 0.3) showed positive values of $\mathrm{\Delta}M$. Using SANSPOL, we observed that the samples with negative $\mathrm{\Delta}M$ values show the presence of a magnetically disordered layer, while the Mn-richest sample did not show any increase of the magnetic radius with the applied magnetic field. Additionally, the study isolated different contributions to the total magnetic anisotropy, highlighting the significant role of shape anisotropy due to the faceted morphology of the NPs. These findings enhance our understanding of the complex interplay between magnetic interactions, surface structure, and anisotropy in mixed Co/Mn ferrites, offering more profound insights into tuning their magnetic properties, which are important for AC biomedical applications, such as cancer treatment by magnetic fluid hyperthermia. 


\begin{acknowledgement}
	
The authors thank ISIS Neutron and Muon Source for the provision of the beamtime (RB2220620-1). The authors acknowledge the assistance provided by the Advanced Multiscale Materials for Key Enabling Technologies project, supported by the Ministry of Education, Youth, and Sports of the Czech Republic. Project No. CZ.02.01.01/00/22\_008/0004558, Co-funded by
the European Union. DZ has been supported by Charles University Research Centre program No. UNCE/24/SCI/010, and MG by the Grant Agency of Charles University: GAUK 267323. Infrastructure obtained thanks to OP VVV "Excellent Research Teams", project no. CZ.02.1.01/0.0/0.0/15\_003/0000417 - CUCAM, was used in this study. Furthermore, we would like to thank Dr. Dopita for providing us access to the SAXS instrument.

\end{acknowledgement}

\begin{suppinfo}
The following files are available free of charge.
\begin{itemize}
  \item pdf: Table with previous magnetization results, Instrumentation part, SANSPOL $I_\mathrm{Q}^{-}$ and $I_\mathrm{Q}^{+}$ scattering cross-sections with the core-shell-surfactant model fit and obtained radial distribution of magnetic scattering length density with summarized results from SANSPOL $I_\mathrm{Q}^{-}$ and $I_\mathrm{Q}^{+}$ scattering cross-sections and nuclear-magnetic interference term fits. HRSTEM and conventional TEM micrographs with the estimation of the aspect ratio. Description of the calculation of disorder energy, surface and shape anisotropy constant.
  
\end{itemize}

\end{suppinfo}

\bibliography{References}
\end{document}


\pagestyle{fancy}
\normalsize
\setcounter{page}{1}
\renewcommand{\thepage}{S\arabic{page}}

\begin{titlepage}
	\maketitle
	\thispagestyle{empty}
	\thispagestyle{fancy}
\end{titlepage}

\makeatother
\renewcommand{\figurename}{Figure S}
\renewcommand{\tablename}{Table S}
\newcommand{\reffig}[1]{\ensuremath{\textbf{Figure S}\textbf{\ref{#1}}}}
\newcommand{\reftab}[1]{\ensuremath{\textbf{Table S}\textbf{\ref{#1}}}}

\justifying
\normalsize
\setcounter{page}{2}
\renewcommand{\thepage}{S\arabic{page}}
\section*{Preliminary Results}

In \reftab{table:Magnetization}, selected results are summarized from our previous studies for better comparison\cite{SannaAngotzi2021}. 

\begin{table}[H]
	\centering
	\begin{tabular}{cccccc}
		\hline
		$\begin{array}{l} \textbf{Parameter} \\\hline \textbf{Sample} \end{array}$ & $T_\mathrm{max}$ & $T_\mathrm{diff}$ & $T_\mathrm{B}$ & $\mu_\mathrm{0}H_\mathrm{C}$ & $M_\mathrm{S}^{\mathrm{300K}}$\\
		& (K) & (K) & (K) & (T) & (Am$^{2}$kg$^{-1}$)  \\
		\hline
		\textbf{Co23Mn77} & 231(5) & 235(5)& 171(3) & 0.65(1) & 74(2)   \\
		\textbf{Co37Mn63} & 259(5) & 280(6) & 194(4) & 1.12(3) & 77(2) \\ 
		\textbf{Co74Mn26}& 282(6) & 314(6) & 197(4) & 1.76(1) & 79(2) \\ 
		\textbf{Co86Mn14}& 315(6) & 328(7) & 234(5) & 1.92(1) & 80(2) \\ 
		\hline
	\end{tabular}
	\caption{Summarized results from our previous study\cite{SannaAngotzi2021}. $T_\mathrm{max}$, $T_\mathrm{diff}$, $T_\mathrm{B}$, $\mu_\mathrm{0}H_\mathrm{C}$, and $M_\mathrm{S}^{\mathrm{300K}}$ represent the maximum, furcation point of the ZFC curve (2\% of difference), blocking temperature, coercive field at 10\,K, and saturation magnetization at the 300\,K, respectively.}
	\label{table:Magnetization} 
\end{table}

\section*{Experimental}
\subsection*{SAXS}
Small-angle X-ray scattering (SAXS) experiments were executed at the Department of Mathematics and Physics of the Charles University in Prague using a Xenocs Xeus 2.0 equipped with Cu and Mo K$_\mathrm{\alpha}$ microfocus X-ray sources, toroidal parallel beam producing X-ray mirrors, two sets of beam collimating scatter-less slits and Dectris Pilatus 200k detector with detector distance of 2.50\,m using both Cu and Mo wavelengths to cover the maximum accessible Q-range. 
Measured 2D intensity profiles were azimuthally integrated. We corrected the 1D SAXS patterns to the capillary/sample thickness and the transmission of samples, and we subtracted solvent and capillary signal. The samples were analyzed as toluene dispersion with a concentration of 3.5\,mg/mL in a borosilicate glass capillary with 1.5\,mm diameter, 80\,mm length, and 0.01\,mm wall thickness.
\subsection*{SANS}
SANS experiments (RB2220620-1)\cite{Mameli2022} were performed at the LARMOR instrument (ISIS, UK) with a horizontally applied magnetic field (perpendicular to the neutron beam) in the range of 9.5\,mT - 1.5\,T (saturation field of 1.5\,T). The neutron beam was polarized by a supermirror polarizer, and incident beam polarization (efficiency of better than 0.95 for neutrons above 2.8\,\AA) was reversed by a radio-frequency flipper (efficiency of 0.99). The samples were analyzed as dispersions in d8-toluene in a quartz disc-shaped "banjo" cell. The nuclear scattering cross-section was obtained from the 20$^{\degree}$ sector integration along the applied magnetic field direction at the saturation field of the NPs. The scattering cross-sections with different incident neutron beam polarization were fully radially averaged for analysis. The nuclear-magnetic interference term was obtained as difference of $I_\mathrm{Q}^{+}- I_\mathrm{Q}^{-}$ 2\,D scattering cross-sections followed by full radial averaging. The data reduction was done in Mantid software\cite{ARNOLD2014}, and the data analysis was performed in a GUI interface available on GitHub\cite{GitHub}.

\subsection*{HRSTEM}
High-resolution scanning TEM (HRSTEM) measurements were carried out on JEOL NEOARM 200 F operating at 200\,kV equipped with Schottky-FEG cathode and CS corrector, respectively. The toluene dispersion of NPs was dropped at the Cu grid with 400 mesh coated by carbon foil. Acquisition of HRSTEM micrographs was made in annular bright (ABF) and dark-field (ADF) mode.

\section*{Results from magnetic SANS}
\begin{figure}[H]
	\begin{center}
		\advance\leftskip-0.4cm
		\advance\rightskip0cm
		\includegraphics[width=0.8\textwidth]{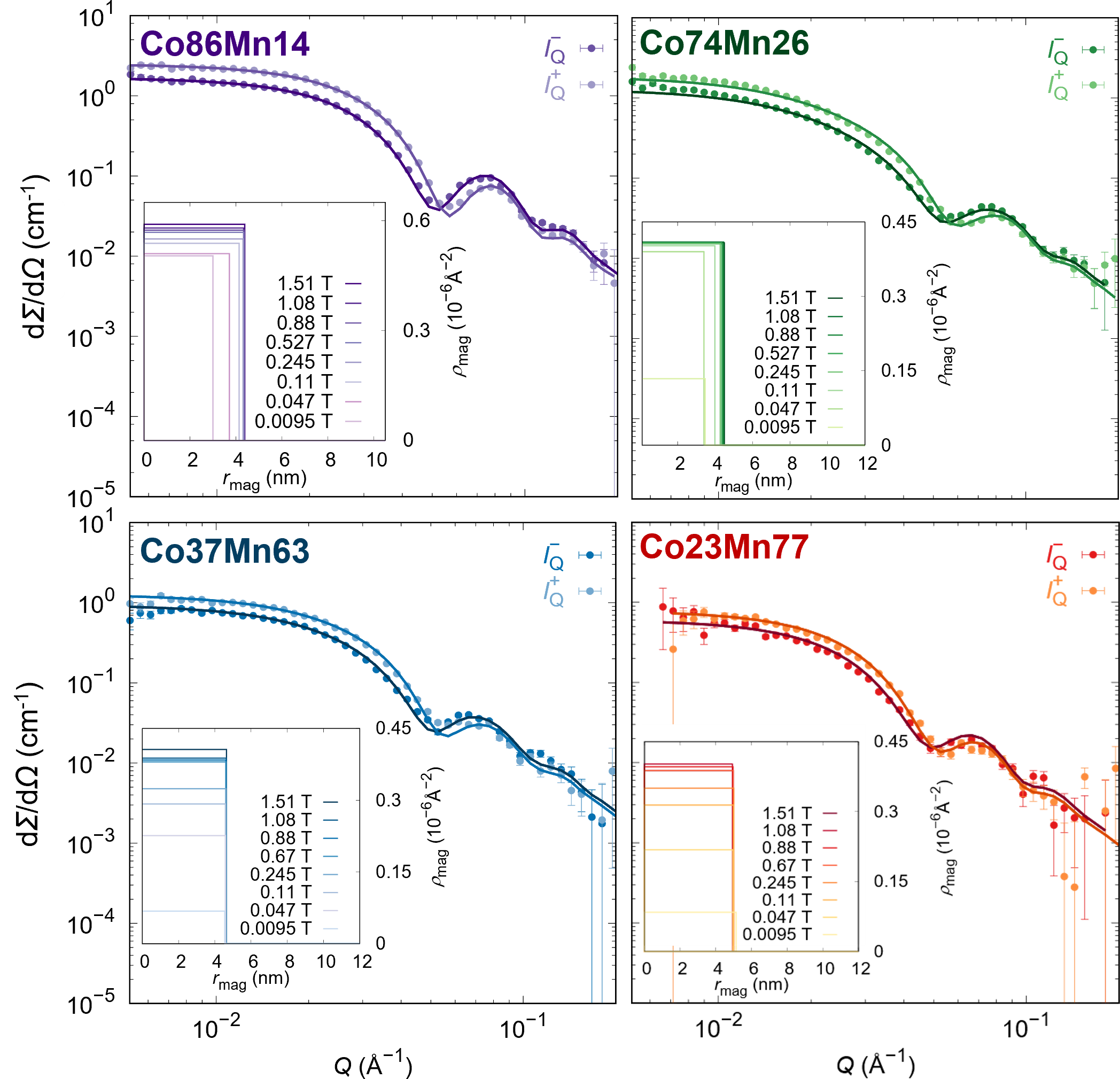}
		\caption{SANSPOL scattering cross sections (points) for the polarization $I_\mathrm{Q}^{-}$ and $I_\mathrm{Q}^{+}$ with core-shell-dead layer form factor refinements (full lines) at 1.51\,T. Insets: Obtained radial distribution of $\rho_\mathrm{mag}$ at different applied magnetic fields.}
		\label{fig:SANSPOL}
	\end{center}
\end{figure}

SANSPOL $I_\mathrm{Q}^{-}$ and $I_\mathrm{Q}^{+}$ scattering cross-sections with the core-shell-surfactant model fit and obtained radial distribution of magnetic scattering length density is presented in \reffig{fig:SANSPOL} with summarized results in \reftab{table:SANSPOL_dispersions}.

\begin{table}[H]
	\centering
	\begin{tabular}{ccccccccc}
		\hline
		$\begin{array}{l} \textbf{Parameter} \\\hline \textbf{Sample} \end{array}$ & $B$ & $r_\mathrm{mag}$ & $d_\mathrm{dis}$ & $\rho_\mathrm{mag}$ & $M_\mathrm{z}$ & $M_\mathrm{ave}$ & $\Delta E_\mathrm{dis}$ & $K_\mathrm{eff-surf}$\\
		& (T) & (nm) & (\AA) & (10$^{-7}$\,\AA$^{-2}$) & (kA/m) & (kA/m) & (10$^{-20}$ J) & (10$^{5}$ Jm$^{-3}$) \\
		\hline
		\multirow{9}{*}{\textbf{Co86Mn14}} 
            & 1.51 & 4.35(2) & 3.5(2) & 5.91(7) & 402(5) & 322(6) & -0.44(5) & 6.1(9) \\
		& 1.08 & 4.37(2) & 3.3(2) & 5.64(6) & 388(4) & 388(7) & -0.10(1) & 4.2(6)\\ 
		& 0.86 & 4.38(2) & 3.2(2) & 5.59(5) & 384(4) & 310(5) & - & -\\ 
		& 0.53 & 4.39(2) & 3.1(2) & 5.46(5) & 376(4) & 306(5) & 0.048(5) & 2.0(3)\\ 
		& 0.25 & 4.40(2) & 3.0(2) & 5.18(7) & 356(5) & 292(6) & 0.051(5) & 0.9(1)\\ 
		& 0.11 & 4.40(4) & 3.0(6) & 4.58(9) & 315(6) & 258(8) & 0.020(2) & 0.36(6)\\ 
		& 0.047 & 4.39(6) & 3.0(6) & 3.4(1) & 232(8) & 189(11) & 0.0042(5) & 0.11(2)\\ 
		& 0.0095 & 4.4(2) & 3(2) & 2.1(2) & 142(12) & 115(15) & - & -\\ 
		\hline
		\multirow{8}{*}{\textbf{Co74Mn26}} 
            & 1.51 & 4.49(2) & 0.068(5) & 4.1(1) & 280(8) & 276(8) & 8.7(2) & 4.2(1)\\ 
		& 1.08 & 4.46(4) & 0.41(3) & 4.1(1) & 281(8) & 274(7) & 6.0(2) & 3.04(8)\\ 
		& 0.86 & 4.41(4) & 0.88(3) & 4.1(1) & 279(7) & 263(7) & 4.5(1) & 2.46(7)\\ 
		& 0.53 & 4.36(4) & 1.41(2) & 4.1(1) & 280(7) & 255(6) & 2.55(7) & 1.48(4)\\ 
		& 0.25 & 4.27(4)  & 2.26(3) & 4.0(1) & 279(7) & 239(6) & 1.04(3) & 0.68(2)\\ 
		& 0.11 & 3.99(4)   & 5.10(3) & 4.0(1) & 276(7) & 193(5) & 0.29(7) & 0.32(8)\\ 
		& 0.047 & 3.41(4) & 10.994(3) &  3.9(1)  & 268(7) & 116(3) & 0.011(3) & -0.0.13(3)\\ 
		& 0.0095 & 3.47(4) & 10.35(4) & 1.4(1) & 93(8) & 43(4) & - & -\\ 
		\hline
		\multirow{8}{*}{\textbf{Co37Mn63}} 
           & 1.51 & 4.64(5) & 3.6(5) & 4.1(1) & 279(7) & 224(9) & 1.1(3) & 4(2)\\ 
		& 1.08 & 4.63(5) & 3.7(5) & 3.9(1) &  267(7) & 212(9) & 0.6(2) & 3(1) \\  
		& 0.86 & 4.61(5) & 3.9(5) & 3.8(1) &  264(7) & 206(9) & 0.4(1) & 2.3(9)\\  
		& 0.66 & 4.63(5) & 3.7(5) & 3.8(1) & 262(7) & 207(9) & 0.4(1) & 1.8(6) \\ 
		& 0.25 & 4.62(6) & 3.8(6) & 3.3(1) & 223(7) & 176(9) & 0.10(3) & 0.5(2)  \\ 
		& 0.11 & 4.58(7) & 4.2(7) & 2.9() & 201(7) & 155(10) & 0.02(6) & 0.2(8) \\ 
		& 0.047 & 4.59(9) & 4.1(9) & 2.3(1) & 156(7) & 121(9) & 0.009(2) & 0.07(3) \\ 
		& 0.0095 & 4.55(4) & 5(4) & 0.7(1) & 47(9) & 35(16) & - & - \\ 
		\hline
		\multirow{8}{*}{\textbf{Co23Mn77}} 
            & 1.51 & 5.20(9) & 2.0(9)  & 3.5(1) & 241(10) & 216(15) & 0.18(4) & 3(1) \\ 
		& 1.08  & 5.18(7) & 2.2(7)  & 3.6(1) & 248(8) & 220(12) & -0.036(8) & 2.7(8)\\ 
		& 0.86  & 5.18(8) & 2.2(8)  & 3.6(1) & 249(9) & 220(13) & -0.059(1) & 2.2(7)\\ 
		& 0.66  & 5.16(9) & 2.4(9)  & 3.6(1) & 247(10) & 216(14) & -0.14(3) & 1.7(5) \\ 
		& 0.25  & 5.18(9) & 2.2(9)  & 3.3(1) & 225(9) & 199(13) & -0.015(3) & 0.6(2)\\ 
		& 0.11  & 5.1(1) & 3(1) & 3.0(1) & 204(9) & 176(13) & -0.037(8) & 0.23(7)\\ 
		& 0.047 & 5.1(1) & 3(1) & 2.1(1) & 142(9) & 119(12) & -0.022(5) & 0.07(5)\\ 
		& 0.0095 & 5.2(4) & 2(4) & 0.8(1) & 56(10) & 50(14) & - & - \\ 
		\hline
	\end{tabular}
	\caption{Summarized results from SANSPOL refinements at different applied magnetic fields, $B$, where $r_\mathrm{mag}$, $d_\mathrm{dis}$, $\rho_\mathrm{mag}$, $M_\mathrm{z}$, $M_\mathrm{ave}$, $\Delta E_\mathrm{dis}$ and $K_\mathrm{eff-surf}$ correspond to the magnetic radius, disorder thickness, magnetic scattering length density of core,  longitudinal magnetization, averaged magnetization, disorder energy and surface contribution to $K_\mathrm{eff}$, respectively.}
	\label{table:SANSPOL_dispersions} 
\end{table}

\newpage
Results from the nuclear-magnetic interference term refined with the core-shell-surfactant model are summarized in 
 \reftab{table:crossterm_dispersions}.

\begin{table}[H]
	\centering
	\begin{tabular}{ccccccccc}
		\hline
	$\begin{array}{l} \textbf{Parameter} \\\hline \textbf{Sample} \end{array}$ & $B$ & $r_\mathrm{mag}$ & $d_\mathrm{dis}$ & $\rho_\mathrm{mag}$ & $M_\mathrm{z}$ & $M_\mathrm{ave}$ & $\Delta E_\mathrm{dis}$ & $K_\mathrm{eff-surf}$\\
		& (T) & (nm) & (\AA) & (10$^{-7}$\,\AA$^{-2}$) & (kA/m) & (kA/m) & (10$^{-20}$ J) & (10$^{5}$ Jm$^{-3}$) \\
		\hline
		\multirow{9}{*}{\textbf{Co86Mn14}} 
            & 1.51 & 4.50(1) & 2.0(1) & 5.27(3) & 362(2) & 318(3) & 12.6(5) & 5.5(3)\\
		& 1.08 & 4.62(1) & 0.8(1) & 4.75(3) & 326(2) & 310(3) & 9.3(4) & 3.5(2) \\ 
		& 0.86 & 4.62(1) & 0.8(1) & 4.74(3) & 326(2) & 309(3) & 7.5(3) & 2.9(2)\\ 
		& 0.53 & 4.61(1) & 0.9(1) & 4.66(3) & 320(2) & 302(3) & 4.4(2) & 1.7(9) \\ 
		& 0.25 & 4.59(1) & 1.1(1) & 4.54(3) & 312(2) & 290(4) & 1.94(7) & 0.76(4) \\ 
		& 0.11 & 4.52(2) & 1.82(62) & 4.18(3) & 287(2) & 255(3) & 0.78(3) & 0.33(2) \\ 
		& 0.047 & 4.04(2) & 6.6(2) & 4.12(6) & 283(4) & 179(4) & 0.166(7) & 0.13(8) \\ 
		& 0.0095 & 3.30(4) & 14.0(4) & 4.0(1) & 275(8) & 95(5) & - & - \\ 
		\hline
		\multirow{8}{*}{\textbf{Co74Mn26}} 
            & 1.51 & 4.49(3) & 0.1(3) & 3.84(8) & 264(6) & 262(8) & 11(1) & 4.0(5) \\ 
		& 1.08 & 4.49(3) & 0.1(3) & 3.77(8) & 259(6) & 257(8) & 7.9(7) & 2.8(3) \\ 
		& 0.86 & 4.45(4) & 0.5(5) & 3.74(1) & 257(5) & 249(8) &  6.20(6) & 2.3(3) \\ 
		& 0.53 & 4.40(4) & 1.0(4) & 3.74(8) & 257(5) & 241(8) & 3.5(3) & 1.4(2) \\ 
		& 0.25 & 4.32(4)  & 1.9(4) & 3.72(9) & 256(6) & 224(8) & 1.5(1) & 0.6(8) \\ 
		& 0.11 & 4.02(4)   & 4.8(4) & 3.72(8) & 256(5) & 182(6) &   0.52(4) & 0.29(3) \\ 
		& 0.047 & 3.43(5) & 10.7(5) &  3.6(1)  & 249(8) & 110(6) & 0.085(8) & 0.12(2) \\ 
		& 0.0095 & 2.84(8) & 16.6(8) & 1.9(1) & 131(9) & 33(4) & - & - \\ 
		\hline
		\multirow{8}{*}{\textbf{Co37Mn63}} 
           & 1.51 & 4.69(2) & 3.15(2) & 3.38(4) & 232(3) & 191(4) & 14(1) & 3.5(5)\\ 
		& 1.08 & 4.60(2) & 4.0(2) & 3.38(4) &  232(3) & 181(4) & 9(1) & 2.5(4)\\  
		& 0.86 & 4.56(3) & 4.4(3) & 3.37(5) &  232(3) & 176(4) & 7.2(2) & 2.0(3) \\  
		& 0.66 & 4.57(2) & 4.3(2) & 3.33(4) & 229(3) & 175(3) & 5.5(6) & 1.5(2) \\ 
		& 0.25 & 4.29(3) & 7.1(3) & 3.31(5) & 227(3) & 144(3)  &  1.6(2) & 0.56(9)\\ 
		& 0.11 & 4.14(3) & 8.6(3) & 3.22(6) & 221(4) & 126(4) &  0.64(7) & 0.26(4)\\ 
		& 0.047 & 3.71(4) & 12.9(4) & 3.16(8) & 217(5) & 86(4) &  0.17(1) & 0.10(2)\\ 
		& 0.0095 & 2.19(8) & 28.1(8) & 3.0(2) & 209(2) & 18(2) & - & - \\
		\hline
		\multirow{8}{*}{\textbf{Co23Mn77}} 
            & 1.51 & 5.25(6) & 1.5(6) & 3.37(9) & 232(6) & 213(10) & 18(5) & 3.5(9) \\ 
		& 1.08  & 5.25(6) & 1.5(6) & 3.43(9) & 236(6) & 217(10) & 13(2) & 2.5(6)\\ 
		& 0.86  & 5.26(8) & 1.4(8) & 3.4(1) & 234(7) & 216(11) & 11(2) & 2.1(5) \\ 
		& 0.66  & 5.25(6) & 1.5(6) & 3.41(9) & 234(6) & 216(9) & 8(1) & 1.6(4)\\ 
		& 0.25  & 5.04(6) & 3.6(6) & 3.36(1) & 231(7) & 187(7) & 2.6(5) & 0.57(1)\\ 
		& 0.11  & 4.82(5) & 5.8(5) & 3.36(9) & 231(6) & 164(7) & 1.1(2) & 0.27(7)\\ 
		& 0.047 & 4.17(8) & 12.3(8) & 3.3(2) & 227(1) & 105(8) & 0.3(5) & 0.11(3)\\ 
		& 0.0095 & 2.6(1) & 28(1) & 3.0(4) & 204(3) & 23(5) & - & -\\ 
		\hline
	\end{tabular}
	\caption{Summarized results from nuclear-magnetic interference term at different applied magnetic fields, $B$, where $r_\mathrm{mag}$, $d_\mathrm{dis}$, $\rho_\mathrm{mag}$, $M_\mathrm{z}$, $M_\mathrm{ave}$, $\Delta E_\mathrm{dis}$ and $K_\mathrm{eff-surf}$ correspond to the magnetic radius, disorder thickness, magnetic scattering length density of core,  longitudinal magnetization, averaged magnetization, disorder energy and surface contribution to $K_\mathrm{eff}$, respectively.}
	\label{table:crossterm_dispersions} 
\end{table}
\newpage
\section*{Disorder energy and surface anisotropy constant}
The disorder energy, $E_\mathrm{dis}$\, is defined as
\begin{equation}
	 E_\mathrm{dis} = \mu\cdot H\cdot M_\mathrm{z}(H)\cdot [V_\mathrm{mag}(H)-V_\mathrm{mag}(H_\mathrm{min})],
\end{equation}
where $M_\mathrm{z}(H)$, $V_\mathrm{mag}(H_\mathrm{max})$ and $V_\mathrm{mag}(H_\mathrm{min})$ are the longitudinal magnetization at the applied field, and magnetized volumes at $H_\mathrm{max}$  and $H_\mathrm{min}$, respectively.
Afterward, they accessed the surface contribution to effective anisotropy constant according to the following equation:
\begin{equation}\label{Keff_s}
    K_\mathrm{eff-surf} = \frac{\partial E_\mathrm{dis}}{\partial V_\mathrm{mag}},
\end{equation}
where $\partial E_\mathrm{dis}$ and $\partial V_\mathrm{mag}$ are the derivative of the disorder energy and of the magnetic volume, respectively. From the effective anisotropy constant, the spatially resolved surface anisotropy constant, $K_\mathrm{S}$, can be obtained:
\begin{equation}\label{Ks}
K_\mathrm{S} = 
K_\mathrm{eff-surf}\cdot\frac{r_\mathrm{mag}}{3}
\end{equation}

\section*{HRSTEM micrographs}
\begin{figure}[H]
	\begin{center}
		\advance\leftskip-0.4cm
		\advance\rightskip0cm
		\includegraphics[width=0.8\textwidth]{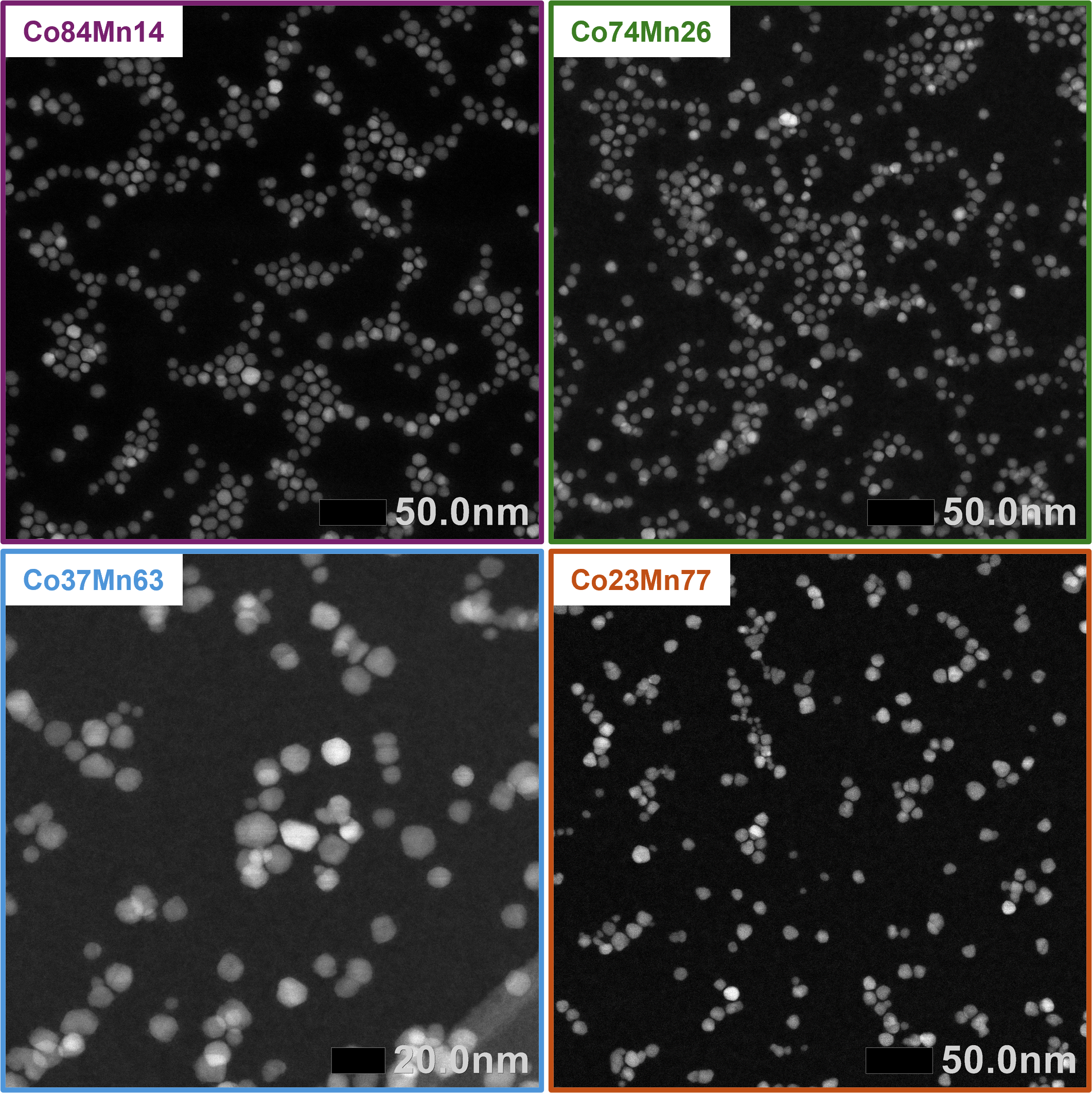}
		\caption{HRSTEM micrographs in ADF mode for the samples Co86Mn14, Co74Mn26, Co37Mn63, and Co23Mn77.}
		\label{fig:HRSTEM}
	\end{center}
\end{figure}

\begin{figure}[H]
	\begin{center}
		\advance\leftskip-0.4cm
		\advance\rightskip0cm
		\includegraphics[width=0.8\textwidth]{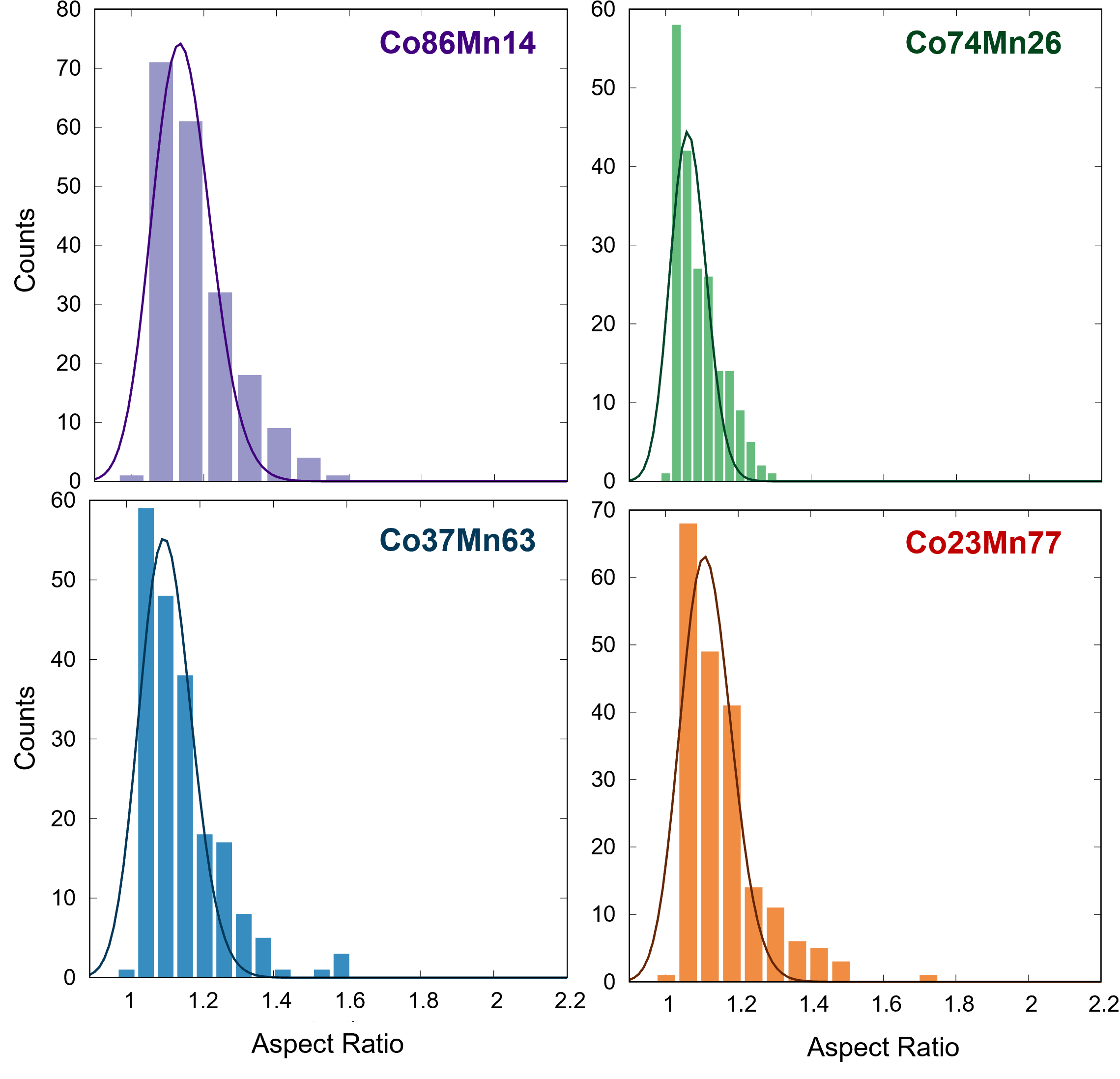}
		\caption{Distribution of aspect ratio of the nanoparticles obtained from TEM micrograph analysis with lognormal distribution function fit.}
		\label{fig:AR}
	\end{center}
\end{figure}

HRSTEM micrographs of all samples in ADF mode are presented in \reffig{fig:HRSTEM} and histogram of aspect ratio with lognormal distribution function in \reffig{fig:AR}.


\begin{figure}[H]
	\begin{center}
		\advance\leftskip-0.4cm
		\advance\rightskip0cm
		\includegraphics[width=0.8\textwidth]{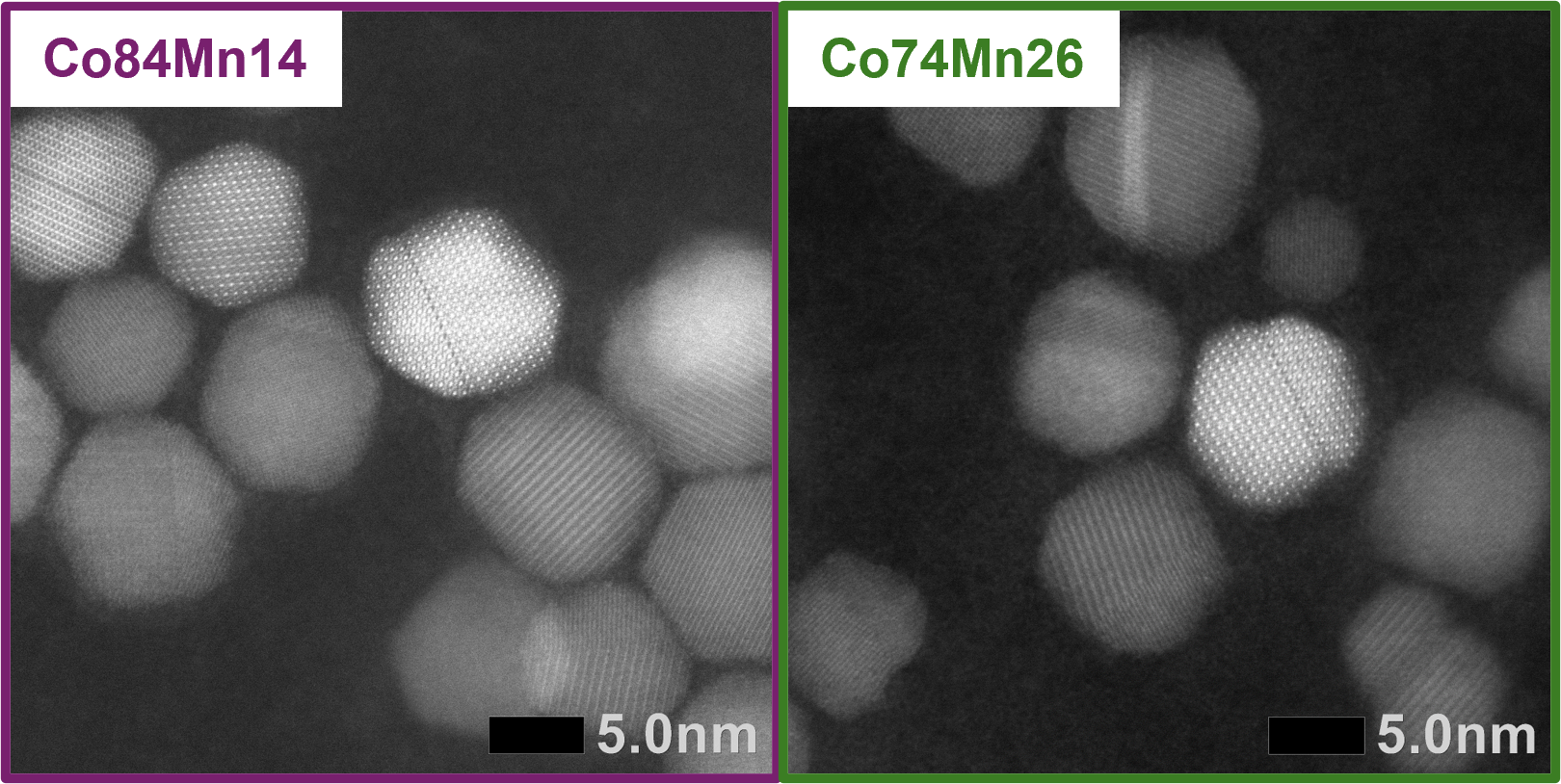}
		\caption{Detail on defects in the crystal structure of Co86Mn14 and Co74Mn26 samples.}
		\label{fig:Defects}
	\end{center}
\end{figure}
\newpage
\subsection*{Shape anisotropy}
The anisotropic shape of particles affects the demagnetizing energy accordingly: 

\begin{equation}
    E_\mathrm{d} = -\frac{1}{2}\mu_\mathrm{0}MH_\mathrm{d}
\end{equation}

with $\mu_\mathrm{0} = 4\pi\cdot 10^{-7}$\,H/m as a permeability of free space, $H_\mathrm{d}$ is a demagnetization field, which depends on the elongation with a symmetry trace: 

\begin{equation}
    E_\mathrm{d} = -\frac{\mu_\mathrm{0}}{2}\begin{pmatrix}
    M_\mathrm{x} & M_\mathrm{y} & M_\mathrm{z}\\
\end{pmatrix}\begin{pmatrix}
    N_\mathrm{xx} & 0 & 0\\
    0 & N_\mathrm{yy} & 0 \\
    0 & 0 & N_\mathrm{zz}\\
\end{pmatrix}\begin{pmatrix}
    M_\mathrm{x}\\
    M_\mathrm{y} \\
    M_\mathrm{z} \\
\end{pmatrix}
\end{equation}

that leads to:

\begin{equation}
    E_\mathrm{d} = -\frac{\mu_\mathrm{0}}{2}(N_\mathrm{xx}M^{2}_\mathrm{x}+N_\mathrm{yy}M^{2}_\mathrm{y}+N_\mathrm{zz}M^{2}_\mathrm{z})
\end{equation}

for uniformly magnetized nanoparticles, the perpendicular components of the demagnetizing energy are the same. If we consider the change in energy due to magnetization rotation in the x-z plane, the z and x components of the magnetization are related by $M^{2}_\mathrm{x} = (1-M_\mathrm{z})^2$ and by applying $M_\mathrm{z} = m_\mathrm{z}M_\mathrm{S}$ the demagnetizing energy has a form of shape anisotropy: 

\begin{equation}
    K_\mathrm{shape} = -\frac{\mu_\mathrm{0}M_\mathrm{S}}{2}(N_\mathrm{xx} - N_\mathrm{zz})m^2_\mathrm{z}
\end{equation}

where for prolate spheroid the $N_\mathrm{xx}$ and $N_\mathrm{zz}$ are given\cite{Fernandez_shape,Moreno_shape}:

\begin{equation}
    N_\mathrm{zz} = \frac{4\pi}{A^2-1}(\frac{A}{\sqrt{A^2-1}}\ln(A+\sqrt{A^2-1})-1)
\end{equation}

\begin{equation}
    N_\mathrm{xx} = \frac{4\pi-N_\mathrm{zz}}{2}
\end{equation}

where A is the aspect ratio.
\newpage
\printbibliography